\documentstyle[12pt,psfig,epsf]{article}
\begin{document}
\begin{titlepage}
\title{
{\bf Curvature representation of the gonihedric action}
}%title ends
{\bf
\author{ 
G.Koutsoumbas\\
Physics Department, National Technical University \\
Zografou Campus, 15780 Athens, Greece\\
\vspace{1cm}\\
G.K.Savvidy\\
The Niels Bohr Institute\\
Blegdamsvej 17, DK-2100 Copenhagen, Denmark\\
\vspace{1cm}\\
K.G.Savvidy\\
Princeton University, Department of Physics\\
P.O.BOX 708, Princeton, New Jersey 08544, USA
}%author ends
}
\date{}%in order NOT to write the date
\maketitle
\begin{abstract}
\noindent

We analyse the curvature representation of the gonihedric 
action $A(M)$  for the cases when the dependence on the 
dihedral angle is arbitrary.

\end{abstract}
\thispagestyle{empty}
\end{titlepage}
\pagestyle{empty}

\section{Introduction}
\vspace{.5cm}

There is a strong belief that QCD is equivalent to a string theory
and that this equivalence could play an 
important role in the development of the theory of strong 
interactions \cite{gross,wilson,thooft,hooft,nilsen}. 
The solution of the two-dimensional QCD in terms of strings 
has been invoked to put forward the ultimate conjecture that QCD is 
actually equivalent to string theory. The $QCD_{2}$ string has the property 
to suppress folding of the surface and therefore has a kind of 
extrinsic curvature \cite{gross}.

In the recent articles \cite{savvidy1,savvidy2,savvidy3}
the authors suggested a 
geometrical string, which directly extends the notion of the 
Feynman integral over paths to an integral over surfaces.
In this theory the surface 
which degenerates into a single particle world line 
has an amplitude which is proportional to the length of 
world line. This approach was motivated by the fact 
that in the first approximation the 
motion of the particle with internal structure 
is described by the action which should be proportional 
to the length of the spacetime path \cite{savvidy3}.
This principle, together with the continuity principle allows
to define the action of the theory $A(M)$ . 
For the smooth surfaces 
this action is proportional to the mean extrinsic curvature 
and has the property to suppress the folding of 
the surface \cite{savvidy3}. The degree of suppression depends
on the gonimetric factor $\Theta$ in the definition of the action
$A(M)$ and can be made arbitrary large. For that the $\Theta$ 
should grow fast enough near the dihedral angles close to $\pi$
which describe a flat surfaces \cite{savvidy3}.

In the subsequent publication \cite{schneider} the authors have 
found that the action $A(M)$ has an equivalent representation in 
terms of the total curvature $k(E)$ of the 
polygons $C_{E}$ which appear in the intersection of the plane $E$ 
with the given surface $M$. The curvature $k(E)$ should then 
be integrated over all planes $E$ intersecting the surface $M$.
This representation has been proven for the special case of the action
$A(M)$, when the action depends linearly on dihedral angle
$\alpha$. To provide necessary rigidity of the polygons $\{C_{E}\}$ 
and the corresponding suppression of the folding of the surface
it is important to have generalization of the 
curvature representation for the cases when the dependence on 
the dihedral angle $\alpha$ is arbitrary \cite{savvidy3}. The aim of 
this article is to perform numerical analyses of the curvature 
representation of the gonihedric action for the cases when dependence 
on the dihedral angle is arbitrary.

\section{The Action}

The gonihedric string has
been formulated in an Euclidean space $R^{d}$, where the
Feynman integral coincides with the partition function for the
randomly fluctuating surfaces. In the continuum Euclidean space $R^{d}$
the random surface is associated with a connected polyhedral
surface $M$ with vertex coordinates
$X_{i}$, where $i = 1,..,\vert M \vert$ and $\vert M \vert$ is the
number of the vertices. The action is defined as \cite{savvidy1}

$$A(M) = \sum_{<ij>} \vert X_{i} - X_{j} \vert
\cdot \Theta(\alpha_{ij}),\eqno(1)$$
where the gonimetric factor $\Theta$ is defined as

$$\Theta(\pi) = 0 ,~~~~~ \Theta(\alpha) \geq 0 ,\eqno(2)$$
$$ \Theta(2\pi -\alpha) = \Theta(\alpha), \eqno(3)$$
the summation is over all edges $<ij>$ of $M$ and $\alpha_{ij}$
is a dihedral angle between two neighbor faces
(flat triangles) of $M$ in $R^{d}$ having a common edge $<ij>$.
The property (3) always allows to consider the argument of the $\Theta$ 
function to be in the interval $(0,\pi)$ in all subsequent 
computations. One can say  that we are measuring the dihedral 
angle $\alpha_{ij}$ from that side of the surface where 
it is less than $\pi$.
As it is easy to see the action indeed has the physical dimension
of the length and the degree of the suppression of the folding of the 
surface depends on $\Theta$ \cite{savvidy3}. In 
cases when the $\Theta$ is equal to 
$$ \Theta = \vert \pi - \alpha \vert^{\varsigma}$$
and $\varsigma < 1$ the suppression can be made arbitrary large.
The physical
properties of this model and possible connection with QCD 
have been discussed in \cite{savvidy1,savvidy2,savvidy3}.

As we stress above the new representation of action $A(M)$ 
have been proven \cite{schneider} for the
special case when the  factor 
$\Theta$ is linear 
$$\Theta(\alpha) =  \vert \pi - \alpha \vert . \eqno(4)$$
In the present article we shall perform the numerical 
analyses of the curvature representation for the cases 
when the dependence on the dihedral 
angle is arbitrary. The numerical integration shows that one can 
always select the dependence from the dihedral angle so as 
to provide necessary rigidity of the polygons $\{ C_{E} \}$  
and therefore corresponding suppression of the folding of 
the surfaces.  This may help to proceed with the analytical 
results for the general case. 

\section{Curvature representation}

If we denote by $\alpha^{E}_{ij}$ the angle which appears
in the intersection of the plane $E$ with dihedral angle
$\alpha_{ij}$ of the edge $<ij>$, then the total 
curvature $k(E)$ of the polygon or polygons appearing in 
the intersection is equal to the following sum 
\cite{schneider}

$$k(E) = \sum_{<ij>} \vert \pi - \alpha^{E}_{ij} \vert.
\eqno(5)$$
Here we imply that $\alpha^{E}_{ij}$ is equal to $\pi$
for the edges $<ij>$ which are not intersected by the 
given plane $E$. With this definitions the gonihedric 
action $A(M)$ is equal to \cite{schneider}

$$A(M) = \int k(E) dE ,\eqno(6)$$
where $dE$ is the invariant measure of two-dimensional 
planes $E$ in $R^{3}$ \cite{santalo,schneider1,ambartzumian}. 
This measure can be defined explicitly 
in $R^{3}$ through the coordinates of the unit vector 
$\vec n(\theta,\phi)$ which is normal to the plane $E$ and 
through the distance $\rho$ of the plane $E$ from the origin
\cite{santalo}

$$dE = \frac{1}{2\pi}d\rho \sin\theta d\theta d\phi . \eqno(7)$$
To prove the representation (6) let us substitute (5) and (7) 
into (6), this yields \cite{schneider}

$$A(M) = \int \sum_{<ij>} 
\vert \pi - \alpha^{E}_{ij} \vert dE
= \sum_{<ij>}
\int \vert \pi - \alpha^{E}_{ij} \vert dE$$
$$= \sum_{<ij>} \frac{1}{2\pi}
\int \vert \pi - \alpha^{E}_{ij} \vert 
\int^{\vert X_{i}- X_{j}\vert \vert \cos \theta \vert }_{0} d\rho 
\sin\theta d\theta d\phi $$
$$= \sum_{<ij>} \vert X_{i} - X_{j} \vert
\cdot <\vert \pi - \alpha^{E}_{ij}\vert >,\eqno(8)$$ 
where

$$<\vert \pi - \alpha^{E}\vert > =
\frac{1}{2\pi}\int 
\vert \pi - \alpha^{E} \vert 
\vert \cos \theta \vert \sin \theta~ d\theta d\phi 
\equiv \Theta(\alpha)  .\eqno (9)$$
Now one should prove, that $\Theta(\alpha)$ in (9),
after averaging over all orientations of the plane $E$
is indeed equal to $\vert \pi - \alpha \vert$ (4). This
has been proven in \cite{schneider}, but here we 
will present a new
proof which allows to apply numerical integration 
to  more complicated cases.

We have the following expression for 
$\vert \pi - \alpha^{E}\vert$

$$\vert \pi - \alpha^{E}\vert = 
\arccos \left(-\frac{\vert \vec{r_{1}} \vec{r_{2}} \vert}
{\vert \vec{r_{1}} \vert \cdot \vert \vec{r_{2}} \vert} \right)$$
$$=\arccos \left(-\frac{\cos(\alpha) + \tan^{2}(\theta) 
\cos(\phi) \cos(\phi - \alpha)}
{ \sqrt{1+\tan^{2}(\theta) \cos^{2}(\phi)} \cdot 
\sqrt{1+\tan^{2}(\theta) \cos^{2}(\phi - \alpha)} }\right), 
\eqno(10)$$
where $\alpha^{E} = \alpha^{E}(\alpha , \theta , \phi)$ is the 
angle in the intersection of the plane $E$ with dihedral angle 
$\alpha$. 

It is very hard to integrate directly 
(9) and (10) to get (4). Instead of that  we would like to 
remark, that from (9) and (10) it follows that

$$\Theta(\pi) =0,~~~ \Theta(0) = \pi \eqno(11)$$
and that 

$$\Theta(\pi - \alpha) = \pi - \Theta(\alpha)  \equiv
<\alpha^{E}>, \eqno(12)$$
where as before we imply that all angles are in the 
interval $(0,\pi)$ (see property (2)).

The basic property of the dihedral angle $\alpha^{E}$ is 
the $additivity$ property

$$\alpha^{E}(\alpha_{1}+\alpha_{2},\theta ,\phi) = 
\alpha^{E}(\alpha_{1},\theta ,\phi) +
\alpha^{E}(\alpha_{2},\theta ,\phi -\alpha_{1}) \eqno(13)$$
which follows from the geometrical picture or one can 
check it directly using explicit formula (10).

Now substituting (13) into (9) and using (12) we will get

$$\Theta(\alpha_{1}+\alpha_{2}) = 
\Theta(\alpha_{1}) + \Theta(\alpha_{2}) - \pi . \eqno(14)$$
The last functional equation together with the boundary 
condition (11) has unique solution

$$\Theta(\alpha) = <\vert \pi - \alpha^{E} \vert >
= \vert \pi - \alpha \vert . \eqno(15)$$
This completes the proof of the curvature representation 
(6) and (5) for $A(M)$ in the particular case (4). 

\section{Extended curvature}

Our aim is to extend this result into the cases when 
$\Theta(\alpha)$ is an arbitrary function of 
$\vert \pi - \alpha \vert$. Generally one can expect that 

$$A(M) = \int k_{\tau}(E) dE = 
\sum_{<ij>} \vert X_{i} - X_{j} \vert
\cdot \Theta(\alpha_{ij}), \eqno(16) $$
where the extended curvature $k_{\tau}(E)$ is defined as

$$
k_{\tau}(E) = \sum_{<ij>}\tau (\vert \pi - \alpha^{E}_{ij}\vert) 
\eqno(17)
$$
and the function ${\Theta}$ is defined by the 
integration over all planes $E$ 

$$\Theta(\alpha)=  \frac{1}{2\pi}\int 
\tau(\vert \pi - \alpha^{E} \vert ) 
\vert \cos(\theta) \vert \sin(\theta) d\theta d\phi . \eqno(18)$$
Now the problem is to
select an appropriate function $\tau$ in the 
definition of the total 
curvature $k_{\tau}(E)$ such that the integration over all planes
$E$ will produce the gonimetric 
factor $\Theta$ with the necessary properties
(2) and (3).

The most important cases are when $\tau$ is a  power of 
$\vert \pi - \alpha^{E} \vert$

$$\tau~~~=~~~\vert \pi - \alpha^{E} \vert^{n},~~~~~~~~n=1,2,... \eqno(19) $$
or a root function
$$\tau~~~=~~~\vert \pi - \alpha^{E} \vert^{1/n},~~~~~~~n=1,2,... \eqno(20) $$
It is also interesting to compute the two angle correlation function

$$\Theta_{nm}(\alpha_{1},\alpha_{2}) = 
<\vert \pi - \alpha^{E}(\alpha_{1}) \vert^{n} \cdot
\vert \pi - \alpha^{E}(\alpha_{2}) \vert ^{m} > \eqno(21)$$
which is defined by the integral

$$\Theta_{nm}(\alpha_{1},\alpha_{2})= \frac{1}{2\pi}\int 
\vert \pi - \alpha^{E}(\alpha_{1},\theta,\phi) \vert ^{n}~
\vert \pi - \alpha^{E}(\alpha_{2},\theta,\phi -\alpha_{1}) 
\vert ^{m}~\vert \cos \theta \vert \sin \theta d\theta d\phi  
\eqno(22)$$
As it is easy to see from (9),(10) and from the last integral 
it is almost impossible to 
compute these explicitly, therefore we should use 
the numerical integration to study their behaviour. 

\section{Numerical results}

In figure 1 we show the result of the integration over all planes 
$\{E\}$ in (18) for various 
 functions $\tau$. As we see the qualitative behaviour of the 
gonimetric function $\Theta$ is very similar to the behaviour of the 
corresponding $\tau$ function. 

\begin{figure}
\centerline{\hbox{\psfig{figure=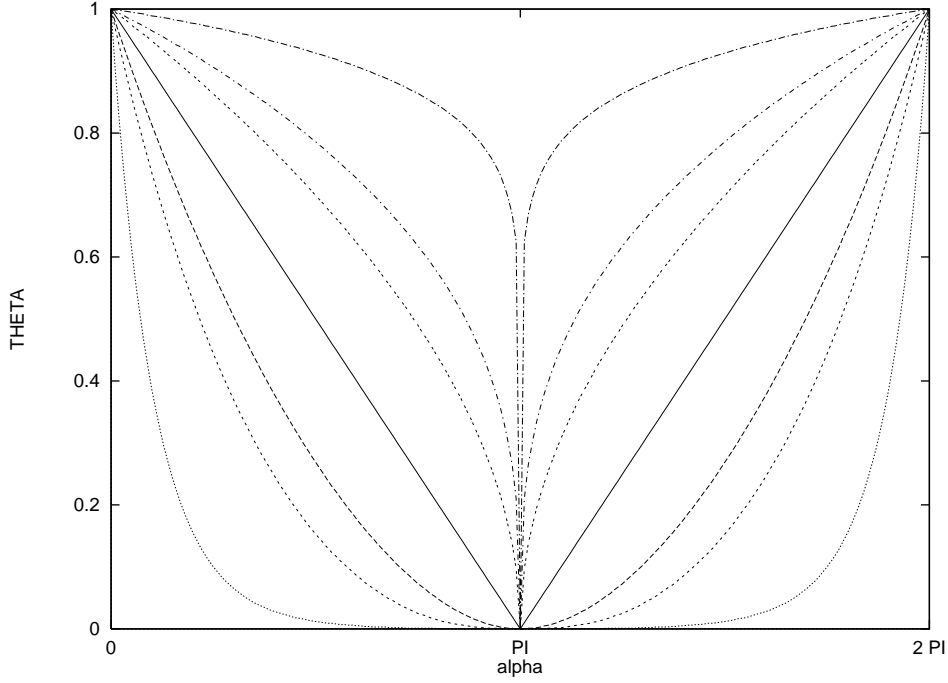,height=9cm,angle=0}}}
\caption[fig1]{Gonimetric function $\Theta$ for the cases when
$\tau$ is given by (19) and (20) and n=1,2,3,10. The lowest curve
corresponds to $n=10$ in (19) and the uppermost curve to $n=10$ in (20).
The curves are normalized to unity at $\alpha=0.$ }
\label{fig1}
\end{figure}

\begin{figure}
\centerline{\hbox{\psfig{figure=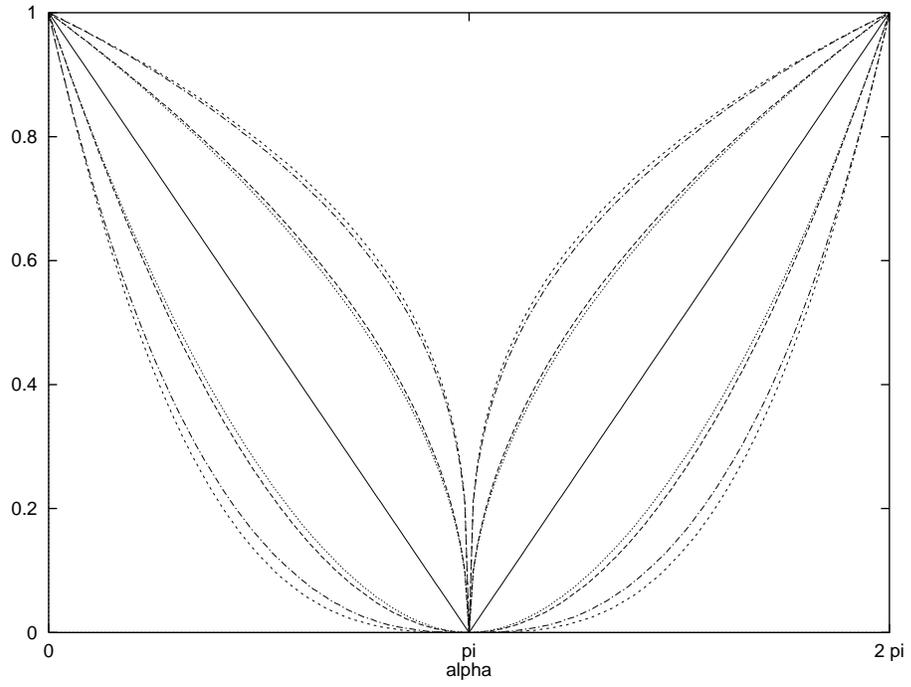,height=9cm,angle=-90}}}
\caption[fig2]{Comparison of the gonimetric function $\Theta$
against $|\pi-\alpha|^n$ for n=1,~2,~3,~1/2,~1/3.}
\label{fig2}
\end{figure}

\begin{figure}
\centerline{\hbox{\psfig{figure=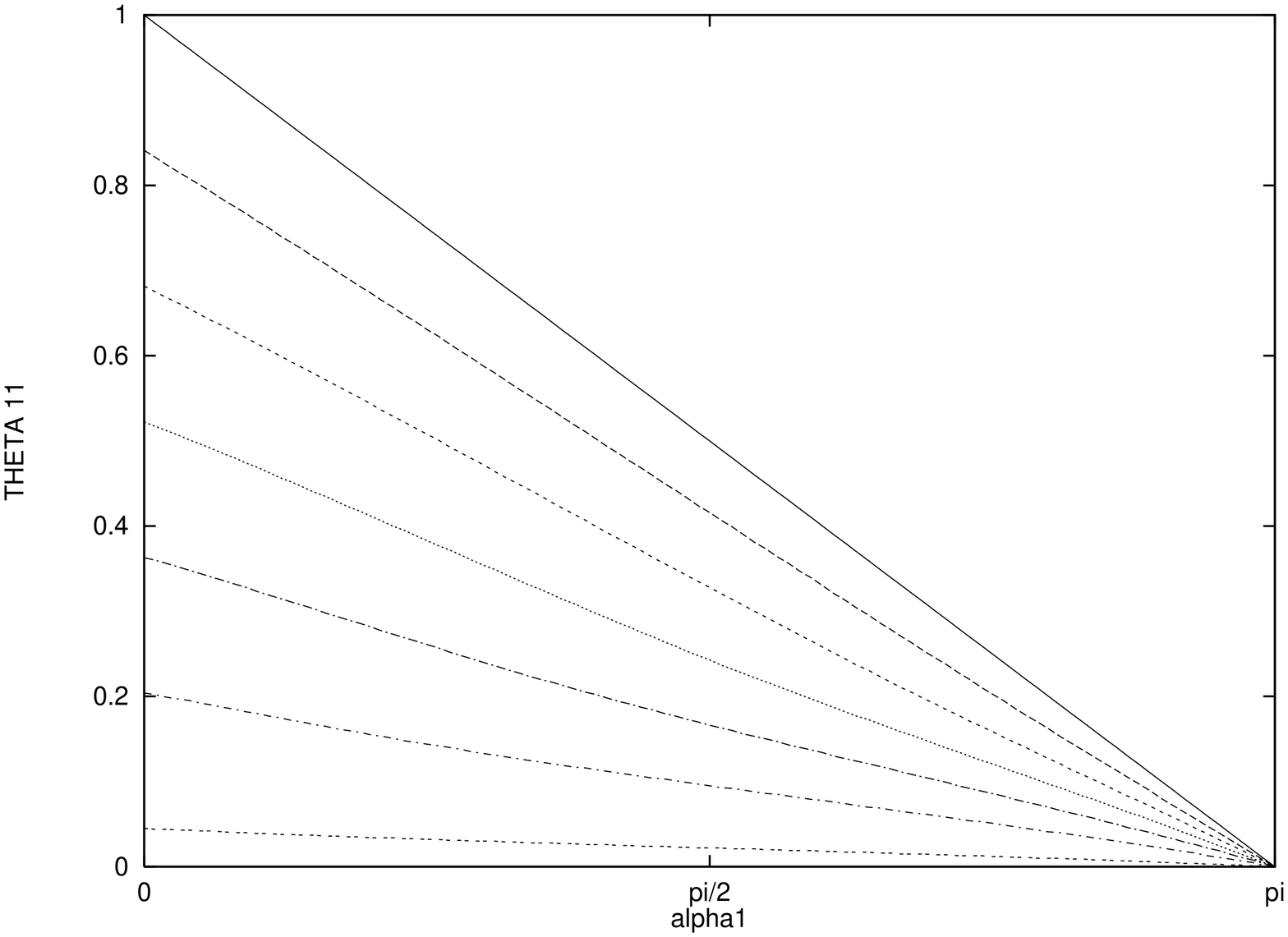,height=6cm,angle=-90}}}
\label{fig31}
\end{figure}

\begin{figure}
\centerline{\hbox{\psfig{figure=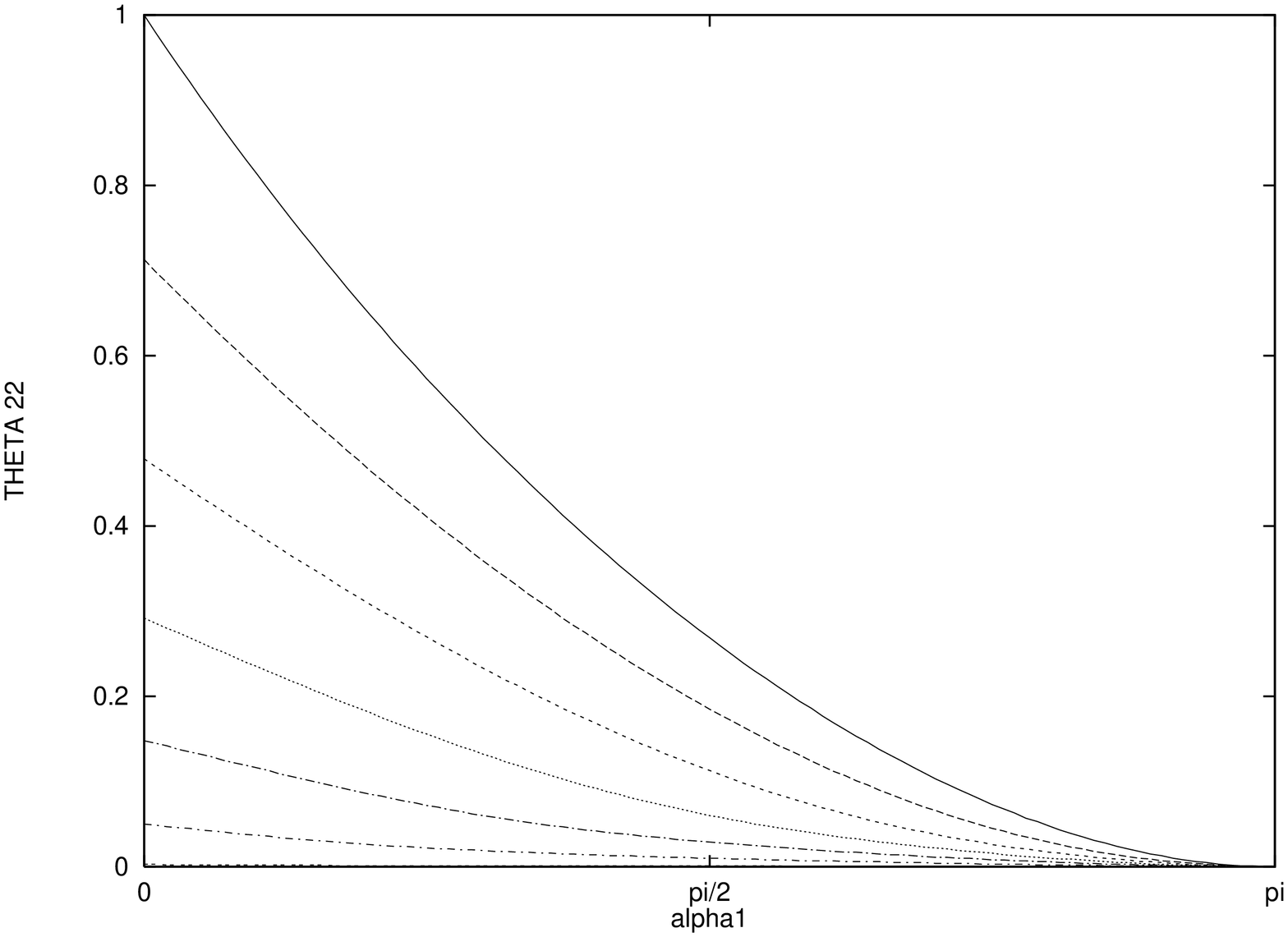,height=6cm,angle=-90}}}
\label{fig32}
\end{figure}

\begin{figure}
\centerline{\hbox{\psfig{figure=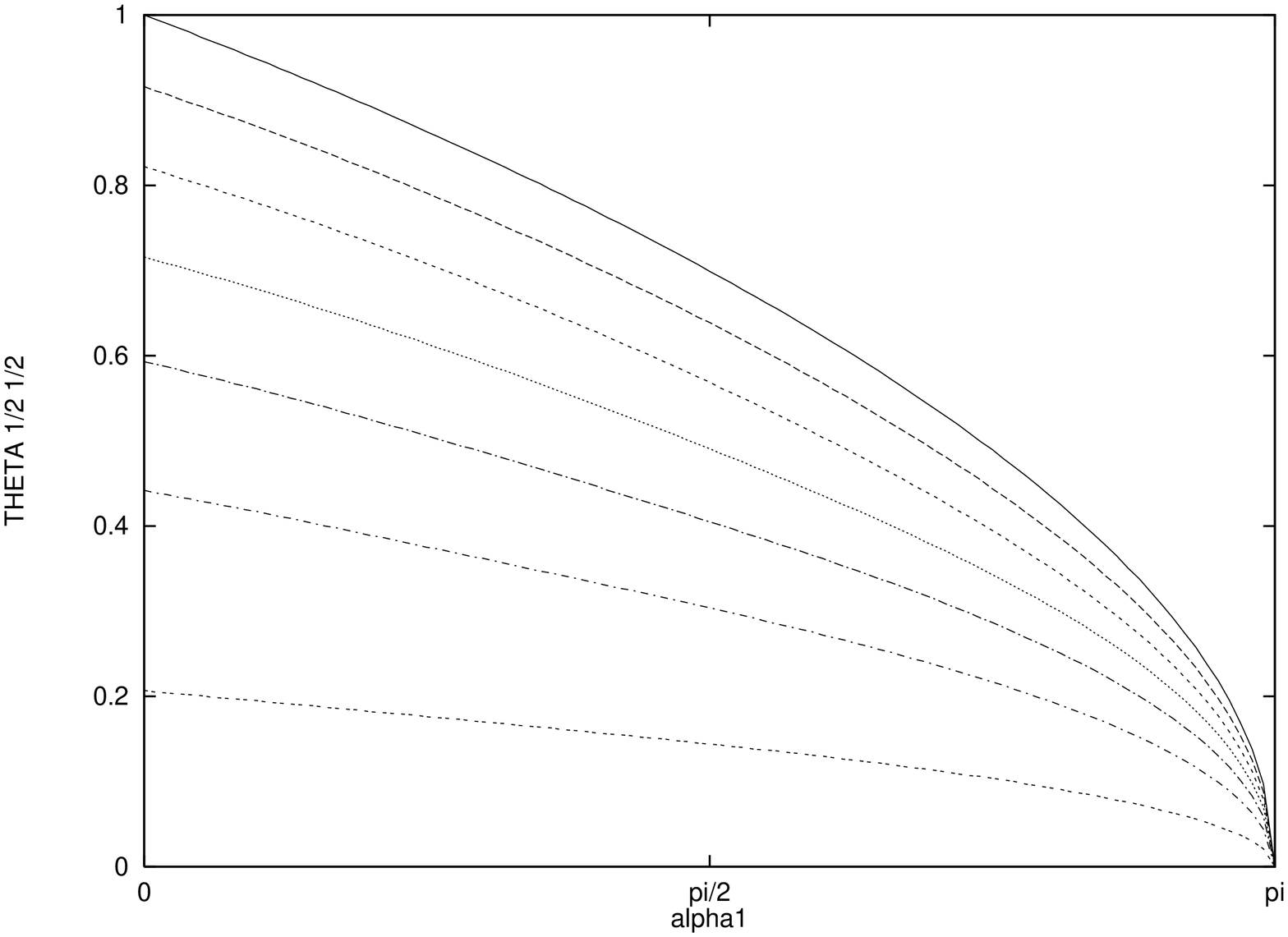,height=6cm,angle=-90}}}
\caption[fig33]{Correlation functions
$\Theta_{nn}(\alpha_1,\alpha_2)$ for 
$\alpha_2=0,~0.5,~1.0,~1.5,~2.0,~2.5,~3.0.$ and 
n=1, n=2 and n=1/2 are shown on figures \ \ref{fig33}(a),
\ \ref{fig33}(b), \ \ref{fig33}(c) respectively. The uppermost curve 
corresponds to $\alpha_2 = 0$ for all three values of n.}
\label{fig33}
\end{figure}

Therefore it is reasonable to ask 
how big is the ``deformation"  of the functions (19) and (20) 
after integration over all planes $\{E\}$ in (18), 
that is the magnitude of the differences
$$ <\vert \pi - \alpha^{E}\vert^{n}>~~~-~~~\vert \pi - \alpha \vert^{n},
\eqno(23) $$
or
$$ <\vert \pi - \alpha^{E}\vert^{1/n}>~~~-~~~\vert \pi - \alpha \vert^{1/n},
\eqno(24) $$
In figure 2 one can see that the differences are not very large, 
but the equality between these two quantities is definitely excluded.
Only for $n=1$ does the equality hold \cite{schneider}.
The question to which we cannot get an answer by numerical analyses 
is how to select the function $\tau$ in (17) to recover the power or
root behaviour of the $\Theta$ function, that is to find out the 
coefficients $c_{k}$ in the $\tau$ series 

$$\tau = \sum_{k} c_{k}~ \vert \pi - \alpha^{E} \vert^{k} \eqno(25)$$
such that 

$$ <\sum_{k} c_{k} \vert \pi - \alpha^{E} 
\vert^{k}>~~~ =~~~\vert \pi - \alpha \vert^{n}. \eqno(26)$$
Our numerical results indicate that the biggest 
coefficient in (26) is  $c_{n}$.

Figure 3 shows the behaviour of the correlation function 
$\Theta_{n,m}$ for various values of the $n$ and $m$. Again 
the function $\Theta_{n,m}$ does not simply coincide with 
$$\vert \pi - \alpha_{1} \vert^{n} \cdot
\vert \pi - \alpha_{2} \vert ^{m}. $$

\section{Comments}

Finally we should stress that the curvature representation 
is very helpful in establishing 
global properties of the action and hence of the partition 
function and the loop Green functions \cite{savvidy3,savvidy4}.

Indeed, with this representation it is easy to prove 
the lower estimate for the action  $A(M)$  \cite{schneider} 
because 
$$\tau = \vert \pi - \alpha^{E} \vert^{1/n} 
\geq \pi^{(1-n)/n}\vert 
\pi - \alpha^{E} \vert ,~~~~~~~~~n \geq 1. \eqno(27)$$
and for the extended curvature (17) we have 

$$k_{\tau}(E) = \sum_{<ij>} \vert \pi -\alpha^{E}_{ij}\vert^{1/n} 
\geq \pi^{(1-n)/n}
\sum_{<ij>} \vert \pi - \alpha^{E}_{ij} \vert
\geq 2\pi^{1/n} , \eqno(28)$$
where we have used the well known inequality for the total curvature
of the polygon \cite{milnor}. For the action this yields

$$A(M) = \int k_{\tau}(E) dE \geq 2\pi^{1/n} \int dE \geq 2\pi^{1/n} 
\Delta ,\eqno(29)$$
where $\Delta$ is the diameter of the surface $M$.  
\vspace{1cm}

{\Large{\bf Acknowledgements}}
\vspace{.5cm}

We would like to thank the Niels Bohr Institute for kind hospitality.
One of the authors (G.K.S.) thanks J.Iliopoulos, N.Papanicolaou,
R.Schneider and 
G.Tiktopoulos for discussions and support.

\vfill

\vfill
\newpage


\begin{thebibliography}{99}

\bibitem{gross} D.Gross and W.Taylor.
Nucl.Phys. B400 (1993) 181 ; B403 (1993) 395

\bibitem{wilson}K.Wilson. Phys.Rev.D10 (1974) 2445 .

\bibitem{thooft}G. 't Hooft. Nucl.Phys. B72 (1974) 461 .

\bibitem{hooft}G. 't Hooft. Nucl.Phys. B75 (1974) 461.

\bibitem{nilsen}H.B.Nielsen and P.Olesen. Phys.Lett. B32 (1970) 203.

\bibitem{savvidy1}R.V. Ambartzumian, G.K. Savvidy, K.G. Savvidy
and G.S. Sukiasian. Phys. Lett. B275 (1992) 99

\bibitem{savvidy2}G.K. Savvidy and K.G. Savvidy. Int. J. Mod. Phys. 
A8 (1993) 3993.

\bibitem{savvidy3}G.K. Savvidy, K.G. Savvidy. Mod.Phys.Lett. 
A8 (1993) 2963.

\bibitem{schneider}G.K.Savvidy and R.Schneider. Comm.Math.Phys. 
161 (1994) 283 

\bibitem{steiner}J.Steiner.Gesammelte Werke.Bond 2.(Berlin,1882) S 171.

\bibitem {minkowski}H.Minkowski.Math.Ann.B57(1903)447.

\bibitem{santalo}L.A.Santalo. Integral Geometry and Geometric 
Probability. (Addison-Wesley, Reading, MA, 1976).

\bibitem{schneider1}R.Schneider and W.Weil. Integralgeometrie.
(Teubner, Stuttgart 1992).

\bibitem{ambartzumian}R.V.Ambartzumian. Combinatorial integral 
geometry. (Wiley,New York, 1987).

\bibitem{milnor}J.W.Milnor.Ann.Math.52(1950)248

\bibitem{savvidy4}G.K. Savvidy and K.G. Savvidy. Mod.Phys.Lett. 
A 11 (1996) 1379 .

\end{thebibliography}
\end{document}